\documentclass{article}

\usepackage{arxiv}
\usepackage{moreverb}
\usepackage{caption}

\usepackage{graphicx}
\usepackage{amsmath}
\usepackage[utf8]{inputenc} % allow utf-8 input
\usepackage[T1]{fontenc}    % use 8-bit T1 fonts
\usepackage{hyperref}       % hyperlinks
\usepackage{url}            % simple URL typesetting
\usepackage{booktabs}       % professional-quality tables
\usepackage{amsfonts}       % blackboard math symbols
\usepackage{nicefrac}       % compact symbols for 1/2, etc.
\usepackage{microtype}      % microtypography
\usepackage{cleveref}       % smart cross-referencing
\usepackage{lipsum}         % Can be removed after putting your text content
\usepackage{graphicx}
\usepackage{doi}

\title{A Bayesian Treatment Selection Design for Phase II Randomised Cancer Clinical Trials}

\date{May 15, 2025}
\newif\ifuniqueAffiliation
\uniqueAffiliationtrue

\ifuniqueAffiliation % Standard variant of author block
\author{ \href{https://orcid.org/0009-0008-7933-6012}{\includegraphics[scale=0.06]{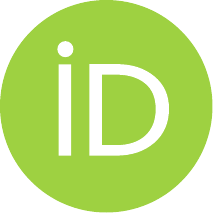}\hspace{1mm}Moka Komaki}\\
	Department of Biostatistics\\
    Devision of Medical Science \\
	Yokohama City University \\
    School of Medicine\\
	Japan\\
	\texttt{mkomaki0603@gmail.com} \\
	\And
	\href{https://orcid.org/0000-0003-0651-438X}{\includegraphics[scale=0.06]{orcid.pdf}\hspace{1mm}Satoru Shinoda} \\
	Department of Biostatistics\\
	Yokohama City University \\
    School of Medicine\\
	Japan\\
    \And
	\href{https://orcid.org/0000-0002-3385-2117}{\includegraphics[scale=0.06]{orcid.pdf}\hspace{1mm}Haiyan Zheng} \\
	Department of Mathmatical Science\\
	University of Bath\\
	The United Kingdom \\
    \And
	\href{https://orcid.org/0000-0003-0696-9659}{\includegraphics[scale=0.06]{orcid.pdf}\hspace{1mm}Kouji Yamamoto} \\
	Department of Biostatistics\\
	Yokohama City University \\
    School of Medicine\\
	Japan\\
}
\else
\usepackage{authblk}

\newbox{\orcid}\sbox{\orcid}{\includegraphics[scale=0.06]{orcid.pdf}} 
\author[1]{%
	\href{https://orcid.org/0009-0008-7933-6012}{\usebox{\orcid}\hspace{1mm}Moka Komaki\thanks{\texttt{mkomaki0603@gmail.com}}}%
}
\author[1,2]{%
	\href{https://orcid.org/0000-0000-0000-0000}{\usebox{\orcid}\hspace{1mm}Elias D.~Striatum\thanks{\texttt{stariate@ee.mount-sheikh.edu}}}%
}
\affil[1]{Department of Biostatistics, Yokohama City University, School of Medicine, Japan}
\affil[2]{Department of Mathmatical Science, University of Bath, The United Kingdom}
\fi

\renewcommand{\shorttitle}{A Bayesian Treatment Selection Design for Phase II Randomised Cancer Clinical Trials}

\hypersetup{
pdftitle={A template for the arxiv style},
pdfsubject={q-bio.NC, q-bio.QM},
pdfauthor={David S.~Hippocampus, Elias D.~Striatum},
pdfkeywords={First keyword, Second keyword, More},
}

\begin{document}
\maketitle

\begin{abstract}
	It is crucial to design Phase II cancer clinical trials that balance the efficiency of treatment selection with clinical practicality. Sargent and Goldberg proposed a frequentist design that allow decision-making even when the primary endpoint is ambiguous. However, frequentist approaches rely on fixed thresholds and long-run frequency properties, which can limit flexibility in practical applications. In contrast, the Bayesian decision rule, based on posterior probabilities, enables transparent decision-making by incorporating prior knowledge and updating beliefs with new data, addressing some of the inherent limitations of frequentist designs. In this study, we propose a novel Bayesian design, allowing selection of a best-performing treatment. Specifically, concerning phase II clinical trials with a binary outcome, our decision rule employs posterior interval probability by integrating the joint distribution over all values, for which the 'success rate' of the bester-performing treatment is greater than that of the other(s).  This design can then determine which a treatment should proceed to the next phase, given predefined decision thresholds. Furthermore, we propose two sample size determination methods to empower such treatment selection designs implemented in a Bayesian framework. Through simulation studies and real-data applications, we demonstrate how this approach can overcome challenges related to sample size constraints in randomised trials. In addition, we present a user-friendly R Shiny application, enabling clinicians to  Bayesian designs. Both our methodology and the software application can advance the design and analysis of clinical trials for evaluating cancer treatments.
\end{abstract}

% keywords can be removed

\section{Introduction}
Phase II clinical trials serve as a foundation for the development of pivotal, large-scape phase III clinical trials. The typical aims include screening new drugs for antitumor activity, exploring new combinations of therapies, and testing new treatment schedules \cite{green}. There is a consensus that randomised controlled trials (RCTs) are the ``gold standard'' to establish early treatment efficacy in phase II clinical trials \cite{Gray}. Because randomisation can balance the prognostic factors (both known and unknown) between treatment groups, providing a proper framework to draw causal inferences \cite{torres}. Conducting RCTs is particularly suitable when the target patient population differs from one that generated existing data. Such situation is of relevance to ethnic differences or the development of molecular targeted therapies \cite{green} \cite{Rub}. Regulatory authorities have also emphasised the importance of RCTs in their guidelines. For instance, the Food and Drug Administration (FDA) issued draft guidance in 2023 on clinical trial considerations to support the accelerated approval of oncology therapeutics, highlighting the potential advantages of RCTs over single-arm trials wherein all patients are treated with the new treatment \cite{FDA}. Similarly, the European Medicines Agency (EMA) has underscored the significance of phase II RCTs \cite{EMA}.

In the design and analysis of phase II clinical trials, Bayesian statistical methods are increasingly utilized due to their ability to incorporate all relevant information, leading to improved parameter estimation. These methods enhance clinical trial efficiency by reducing trial duration, minimizing participant burden, optimizing sample size requirements, and leveraging prior information for more effective data utilization \cite{fors}\cite{Mo}. This is particularly desirable for clinical trials that have limited information to generate \cite{Berry}. As a result, Bayesian statistical designs are gradually being adopted, particularly in early phases of drug development, such as phase II clinical trials \cite{fors}\cite{chen} \cite{yang}. Moreover, Bayesian designs enable coherent inference and decision-making using probabilities, making interpretation easier \cite{te}. As a result, Bayesian designs are currently significant topics of interest in phase II cancer clinical trials, reflecting their potential to enhance the development of effective cancer therapies. 

Sargent and Goldbergs's design (hereafter referred to as SG design)  \cite{sar} \cite{let} is a type of randomised treatment selection design, offering flexibility by allowing the consideration of other factors when the primary endpoint alone is insufficient to determine whether a treatment should proceed to phase III clinical trials. Generally, when a randomized design is conducted with proper error control, a sample size of several hundred patients may be required in phase II RCTs \cite{torres}, which may sometimes not be feasible when developing drugs . In practice, there are also randomized phase II trials conducted with a total sample size of fewer than 100 patients \cite{wild}. Given the growing demand for the use of randomisation, Bayesian designs permitting the selection of a treatment that has a higher efficacy and/or good practicality offer a promissing solution for addressing clinical needs \cite{te}.

When the drug cannot be selected based on the primary endpoint, the decision yielded by the SG design will be made using secondary factors - including toxicity, cost, ease of administration, quality of life, etc. - which are also recommended for consideration in the guidelines \cite{guide}. In some cases, a treatment with a slightly lower response rate may be preferred over one with a higher response rate due to other factors. This design reflects the clinical reality that the success probability of a treatment is only one of many considerations when recommending a treatment for a particular patient. One key feature of the SG design is the inclusion of an ambiguity probability, which allows for slight relaxation of the decision criterion focused on superiority alone. This design has been employed in various oncology studies, including the trial on pertuzumab and trastuzumab with or without metronomic chemotherapy for older patients with HER2-positive metastatic breast cancer (EORTC 75111-10114) \cite{wild}, the phase II RCTs of gefitinib or placebo in combination with tamoxifen in patients with hormone receptor-positive metastatic breast cancer \cite{Osb}, and the randomised phase II study for advanced endometrial carcinoma in the Adjuvant Chemotherapy for Endometrial Cancer (ACE) trial \cite{Egawa}. The SG design is not intended to allow for interim monitoring. The reason is that this design is used in cases where a significant difference in the primary endpoint is not observed. In addition, when drug selection cannot be made based on the primary endpoint, the design relies on secondary factors for treatment selection, which requires following up with all patients until the last one is evaluated. The present work hence does not consider interim adaptation in the selection design. Instead, we aim to develop a Bayesian analogue of treatment selection designs using the SG design as a primary example.

While phase II trials are primarily small-scale exploratory studies, SG designs (RCTs) typically require large sample sizes due to the inclusion of a control group, raising feasibility concerns. Replacing the control group with historical cohort data could introduce risks such as selection bias, observation bias, and confounding bias. Therefore, RCTs that account for sample size constraints are generally preferred. In practice, Bayesian designs have limited uptake due to low awareness among clinicians. A recent survey by the Drug Information Association Bayesian Scientific Working Group (DIA BSWG) uncovered that, insufficient knowledge of Bayesian approaches is perceived as the greatest barrier to implementing Bayesian methods \cite{mu}.

To address these issues, we propose a Bayesian treatment selection design in Section 2 that, compared to the SG design, reduces the required sample size by leveraging prior information. A Bayesian approach to the sample size calculation, drawing inspiration from the SG design, is presented in Section 3. To improve the uptake of our Bayesian design in practice, we develop a web application using R Shiny. We expect this would particularly benefit practioners like clinicians for designing their phase II RCTs. Our proposed methodology is thoroughly evaluated using a simulation study in Section 4 and a case study in Section 5. We then conclude with a discussion in Section 7.

\section{A NOVEL BAYESIAN TREATMENT SELECTION DESIGN}

Consider phase II RCTs comparing two treatments, labelled $A$ and $B$, where the primary endpoint is a binary outcome, i.e., response or no response. We are interested in estimating the response rates. We shall consider Bayesian methods for updating our beliefs based on data generated from the phase II RCT.

Let the number of patients in each treatment group be $n_i$ for $i=A$ or $B$. The sequence of patient responses to $A$ and $B$ will be denoted by $Y_{i1}, Y_{i2}, \ldots, Y_{in_i}$, with each taking the value of $0$ (no response) or $1$ (response). The total number of responses out of  $n_i$ patients is thus $S_i = Y_{i1} + \cdots + Y_{in_i}$. The number of no responses out of $n_i$ patients is $n_i-S_i$. Without loss of generality, let $\pi_A$ denote the response rate on the better treatment (say, $A$) and $\pi_B$ denote the response rate on the poorer treatment (say, $B$). 

In Bayesian inference, $\pi_A$ and $\pi_B$ have a prior distribution each. We assume specifically
\begin{align*}
\pi_i \sim \text{Beta} (\alpha_i, \beta_i)  \ \text{for} \ i = A, B.
\end{align*}
The posterior distribution for the true response rates can then be expressed as 
\begin{align*}
\pi_i|S_i, n_i \sim \text{Beta}(\alpha_i+S_i, \beta_i+n_i-S_i) \ \text{for} \ i = A, B.
\end{align*}

The SG design makes decisions based on the difference observed between $\pi_A$ and $\pi_B$ in a clinical trial. In contrast, sample size determination is based on key metrics such as the probability of correct selection ($P_{Corr}$) and the probability of ambiguity ($P_{Amb}$). A Bayesian analogue to this sample size determination can be constructed by applying Beta posterior distributions to define $P_{Corr}$ and $P_{Amb}$. In what follows we integrate these criteria into the decision-making process to establish a Bayesian framework.

Drawing inspiration from Thall and Simon\cite{Thalla}, we give the probability of correct selection, denoted by $P^*_{Corr}$, as the posterior probability of the difference in response rates being greater than $d$. Here, $d$ is a known quantity representing the clinically meaningful difference. Mathematically,
\begin{align}
P^*_{Corr} &= \mathrm{Pr}[\pi_A-\pi_B > d \ | S_A, S_B, n_A, n_B] \notag \\
&= \int_0^{1-d} \{1-F(\pi+d; \alpha_A+S_A, \beta_A+n_A-S_A)\}f(\pi;\alpha_B+S_B, \beta_B+n_B-S_B) d\pi,
\end{align}
where $f(\cdot ; \alpha, \beta)$ and $F(\cdot; \alpha, \beta)$ are the probability density function and cumulative distribution function of a $\text{Beta} (\alpha, \beta)$ distribution, respectively \cite{Ying}.

The probability of ambiguous selection also following SG design, denoted as $P^*_{Amb}$, is defined as the probability that the difference in true response rates $\pi_A-\pi_B$ is within $|d|$. It is expressed as follows:
\begin{align}
P^*_{Amb} &= \mathrm{Pr}[-d \leq \pi_A-\pi_B \leq d \ | S_A, S_B, n_A, n_B] \notag \\
&= \int_0^{1+d} \{1-F(\pi-d; \alpha_A+S_A, \beta_A+n_A-S_A)\}f(\pi;\alpha_B+S_B, \beta_B+n_B-S_B) d\pi \notag \\
&\quad -\int_0^{1-d} \{1-F(\pi+d; \alpha_A+S_A, \beta_A+n_A-S_A)\}f(\pi;\alpha_B+S_B, \beta_B+n_B-S_B) d\pi.
\end{align}

Let $\lambda^*=P^*_{Corr}+\rho P^*_{Amb} \ (\text{where}\ 0 \leq \rho <1)$ and $\rho$ is selected appropriately for context. When comparing two treatment groups, $\rho$ was set to be $\frac{1}{2}$ in the original SG design. In brevity, the decision rule considered for this design is defined as follows:
\begin{align*}
\begin{cases}
\text{  Choose treatment $A$ as optimal if } \lambda^* > \theta, \\
\text{  Consider other factors for choosing treatment $A$ or $B$, whichever has better behaviour,  if } \lambda^* \leq \theta.
\end{cases}
\end{align*} 
Applying this decision rule with $\rho=1$, it is equivalent to $\lambda^\ast = \mathrm{Pr} (\pi_A - \pi_B \geq -d)$, which may form a Bayesian non-inferiority design for a given margin of $d$. Here, $0<\theta<1$ is the threshold which need to be determined in advance. Furthermore, $\theta$ may be set to a large fraction like 0.80 or 0.90 so a treatment is selected based on compelling evidence of clinical benefit or acceptability in favour of the decision.

\section{THE SAMPLE SIZE CALCULATION}
We now demonstrate the Bayesian sample size determination for the setting considered, wherein a treatment $A$ or $B$ is selected based on $\lambda^*$ defined in Section 2. When concerned with $\rho = 0$, the sample size required is found to ensure the probability of correct selection of a treatment that has superior efficacy than the alternative. 

Aligned with the proposal by Thall and Simon \cite{Thalla} \cite{Thallb}.  
, we proposed the following that requires the expected response rates $\tilde{\pi}_i$ for each group. We assume $\tilde{n}_A = \tilde{n}_B = \tilde{n}$, where $\tilde{n}$ is a positive integer. Let the number of responders be $\tilde{x}_i = \tilde{n} \times \tilde{\pi}_i$. Since the number of responders is likewise an integer, $\tilde{x}_i$ is rounded up to the nearest integer. For example, when $\tilde{n}=30$ and $\tilde{\pi}_A = 0.25$, $\tilde{x}_A$ is calculated to be $\tilde{x}_A=8$. 

The posterior distributions for correct selection and ambiguity are derived from E.q. (1) and E.q. (2) and formulated accordingly. Subsequently, $\lambda^*$ is computed based on this distribution. The minimum sample size $\tilde{n}_{\text{min}}$ is defined as the smallest $\tilde{n}$ such that $\lambda^* > \gamma^*$ always holds, where $\gamma^*$ is a pre-specified threshold. Using the algorithm described above, the results of sample size for each group based on $\lambda^*$ are as follows. Table 1 gives the sample size per treatment group for various configurations of ($\tilde{\pi}_A, \tilde{\pi}_B$) used for generating the new trial data. For illustration, we set $d=0.05$ and $\gamma^*=0.80$ or $\gamma^*=0.90$, assuming $\rho=0$ or $\rho=\frac{1}{2}$ and the use of vague and informative priors, respectively.

\begin{table}[hbtp]
\caption{Determination of a minimum sample size $\tilde{n}_{\text{min}}$ based on $\lambda^*$ for various $\tilde{\pi}_A$ and $\tilde{\pi}_B$, assuming vague and informative priors.}
\label{table:data_type}
\centering
\resizebox{\textwidth}{!}{%
\begin{tabular}{ccccccccc}
\hline
\multicolumn{3}{c}{}&\multicolumn{2}{c}{Response rates}& \multicolumn{2}{c}{$\rho=0$} & \multicolumn{2}{c}{$\rho=\frac{1}{2}$}  \\
\cline{1-9}
&$\text{Beta}(\alpha_A, \beta_A)$&$\text{Beta}(\alpha_B, \beta_B)$&$\tilde{\pi}_A$&$\tilde{\pi}_B$&$\gamma^*=0.90$ & $\gamma^*=0.80$ & $\gamma^*=0.90$ & $\gamma^*=0.80$ \\
\hline
Vague prior&(1, 1)&(1, 1)&0.20 & 0.05 & 53 & 33 & 33 & 13\\
&&&0.25 & 0.10 & 67 & 30 & 38 & 19 \\
&&&0.30 & 0.15 & 72 & 39 & 39 & 19 \\
&&&0.35 & 0.20 & 79 & 39 & 45 & 19 \\
&&&0.40 & 0.25 & 87  & 47  & 52  & 17 \\
&&&0.45 & 0.30 & 93 & 46 & 53 & 26 \\
&&&0.50 & 0.35 & 94 & 54 & 54 & 26 \\
\hline
Informative prior&(2, 8)&(1, 9)&0.20 & 0.05 & 38 & 18 & 18 & 13\\
(Incorporate appropriate prior &(3, 7)&(1, 9)&0.25 & 0.10 & 30 & 10-** & 11 & 10-** \\
information for each group &(3, 7)&(2, 8)&0.30 & 0.15 & 65 & 32 & 39 & 12 \\
of 10 participants.)&(4, 6)&(2, 8)&0.35 & 0.20 & 50 & 19 & 25 & 10-** \\
&(4, 6)&(3, 7)&0.40 & 0.25 & 87  & 39  & 47  & 12 \\
&(5, 5)&(3, 7)&0.45 & 0.30 & 66 & 26 & 33 & 10-* \\
&(5, 5)&(4, 6)&0.50 & 0.35 & 94 & 46 & 54 & 18 \\
\hline
\end{tabular}%一度閾値を超えればいいのか？常に超えている状態じゃないといけないのか
}
\\
  \vspace{0.5\baselineskip} 
\raggedright **an underscore for numbers less than 10.

\end{table}

The sample size required for case of $\rho=\frac{1}{2}$, allowing ambiguous probability for selecting a treatment based on additional factors, is generally smaller compared to that of $\rho=0$. As the threshold $\gamma^*$ increases, the required sample size always increases. This is unsurprising, because a larger threshold $\gamma^\ast$ suggests a higher requirement for decision accuracy. Furthermore, as $\tilde{\pi}_A$ approaches $0.50$, the required sample size increases when the relative difference in the response rates implied by data remains 0.15 in all scenarios (rows of Table 1). The explanation is that the uncertainty in the likelihood increases accordingly with $\tilde{\pi}_i$ getting closer to 0.50. It was suggested that, in most cases, incorporating appropriate prior information reduces the required sample size compared to that of cases using a vague prior. Figure 1 visualises the relationship between sample size and $\lambda^*$, under the scenario of $\tilde{\pi}_A=0.30, \tilde{\pi}_B=0.15, d=0.05$ and $\rho=0$ with a vague or informative prior. As the proposed method shows that the value of $\lambda^*$ is not stable when the sample size is small. In this illustration, the required sample size is defined as the smallest sample size that ensures the resulting $\lambda^\ast$ exceeding $\gamma^*$.

\begin{figure}[h!]
    \centering
	\caption{Sample size determination for the Bayesian approach to selecting a treatment based on $\lambda^\ast$, assuming $\tilde{\pi}_A=0.30$, $\tilde{\pi}_B=0.15$, $d=0.05$, $\rho = 0$ with a vague or informative prior.}
    \includegraphics[width=0.8\textwidth]{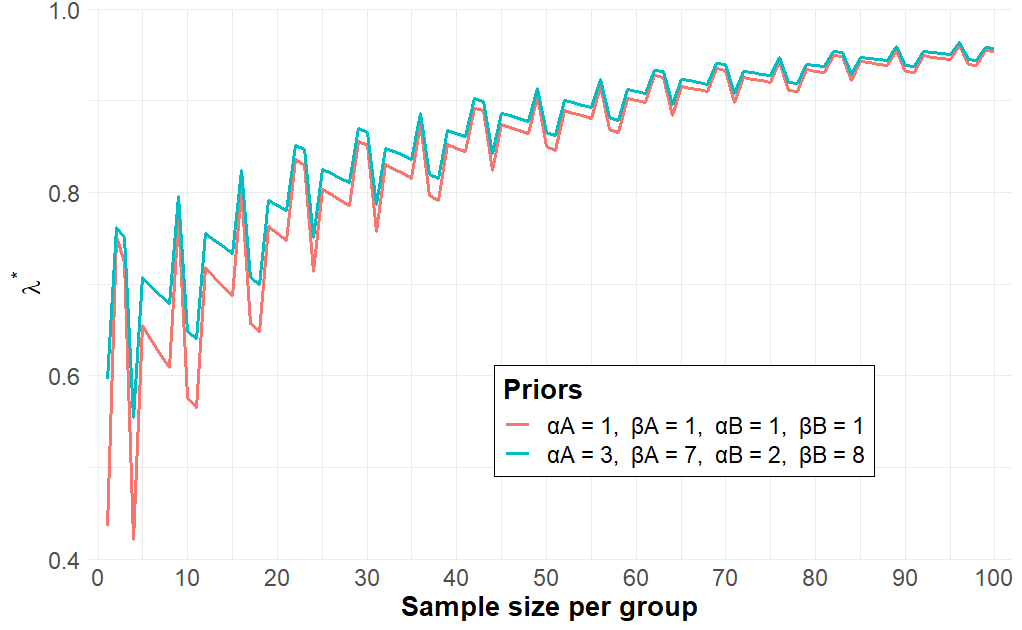}
    \label{fig:sample}
\end{figure}

In the sample size determined based on $\lambda^*$, as shown in Figure 1, its instability associated with small sample sizes suggests potential limitations. To address this, we propose an approach that treats $\tilde{x}_i$ as a random variable rather than deterministically fixing it based on the expected belief response rate, thereby incorporating uncertainty into the sample size design through simulations.

Here, let $\lambda^*_j$ represent the $j$-th value of $\lambda^*$. Each $x^*_i$ is randomly sampled from a binomial distribution, 
\begin{align*}
x^*_i \sim \text{Binomial}(\tilde{n}, \tilde{\pi}_i),
\end{align*}
and $\lambda^*_j$ is calculated $m$ times, corresponding to the number of simulations. We define:
\begin{align*}
\bar{\lambda}^*=\frac{\sum_{j=1}^{m}\lambda^*_j}{m},
\end{align*}
and the minimum sample size required is the point where $\bar{\lambda}^*$ exceeds the threshold $\gamma^*$. Table 2 shows the sample size based on $\bar{\lambda}^*$. Obviously, the sample sizes are larger compared to be those based on $\lambda^*$. Figure 2 visualises the sample size determination concerning $\bar{\lambda^*}$, assuming $\tilde{\pi}_A=0.30, \tilde{\pi}_B=0.15, d=0.05, \rho=0$ and $m=100,000$, based on $\bar{\lambda}^*$, showing more stable values and smoother curves, as compared to Figure 1. 

\begin{table}[hbtp]
\caption{Required sample size based on $\bar{\lambda}^*$ for various $\tilde{\pi}_A$ and $\gamma^*$, assuming $d=0.05$, $\rho=0$ or $\rho=\frac{1}{2}$ and $m=100,000$ with vague and informative prior.}
\label{table:data_type}
\centering
\resizebox{\textwidth}{!}{%
\begin{tabular}{ccccccccc}
\hline
\multicolumn{3}{c}{}&\multicolumn{2}{c}{Response rates}& \multicolumn{2}{c}{$\rho=0$} & \multicolumn{2}{c}{$\rho=\frac{1}{2}$}  \\
\cline{1-9}
&$\text{Beta}(\alpha_A, \beta_A)$&$\text{Beta}(\alpha_B, \beta_B)$&$\tilde{\pi}_A$&$\tilde{\pi}_B$&$\gamma^*=0.90$ & $\gamma^*=0.80$ & $\gamma^*=0.90$ & $\gamma^*=0.80$ \\
\hline
Vague prior&(1, 1)&(1, 1)&0.20 & 0.05 &71 &34 &40 &17 \\
&&&0.25 & 0.10 &94 &43 &52 &21 \\
&&&0.30 & 0.15 &115 &50 &65 &25 \\
&&&0.35 & 0.20 &131 &59 &72 &28 \\
&&&0.40 & 0.25 &145 &64 &79 &31 \\
&&&0.45 & 0.30 &155 &68 &85 &33 \\
&&&0.50 & 0.35 &161 &71 &90 &34 \\
\hline
Informative prior&(2, 8)&(1, 9)&0.20 & 0.05 &60 &24 &30 &10-** \\
(Incorporate appropriate prior&(3, 7)&(1, 9)&0.25 & 0.10 &63 &10-** &22 &10-** \\
information for each group &(3, 7)&(2, 8)&0.30 & 0.15 &106 &43 &54 &15 \\
of 10 participants.)&(4, 6)&(2, 8)&0.35 & 0.20 &102 &26 &45 &10-** \\
&(4, 6)&(3, 7)&0.40 & 0.25 &135 &37 &71 &21 \\
&(5, 5)&(3, 7)&0.45 & 0.30 &125 &37 &58 &10- \\
&(5, 5)&(4, 6)&0.50 & 0.35 &153 &62 &80 &25 \\
\hline
\end{tabular}%一度閾値を超えればいいのか？常に超えている状態じゃないといけないのか
}
\\
  \vspace{0.5\baselineskip} 
\raggedright **an underscore for numbers less than 10.
\end{table}
\begin{figure}[h!]
    \centering
    \caption{Sample size and $\bar{\lambda^*}$ considering the validation of Bayesian approach for a two-arm trial, assuming $\tilde{\pi}_A=0.30, \tilde{\pi}_B=0.15, d=0.05, \rho=0$ and $m=100,000$}
    \includegraphics[width=0.8\textwidth]{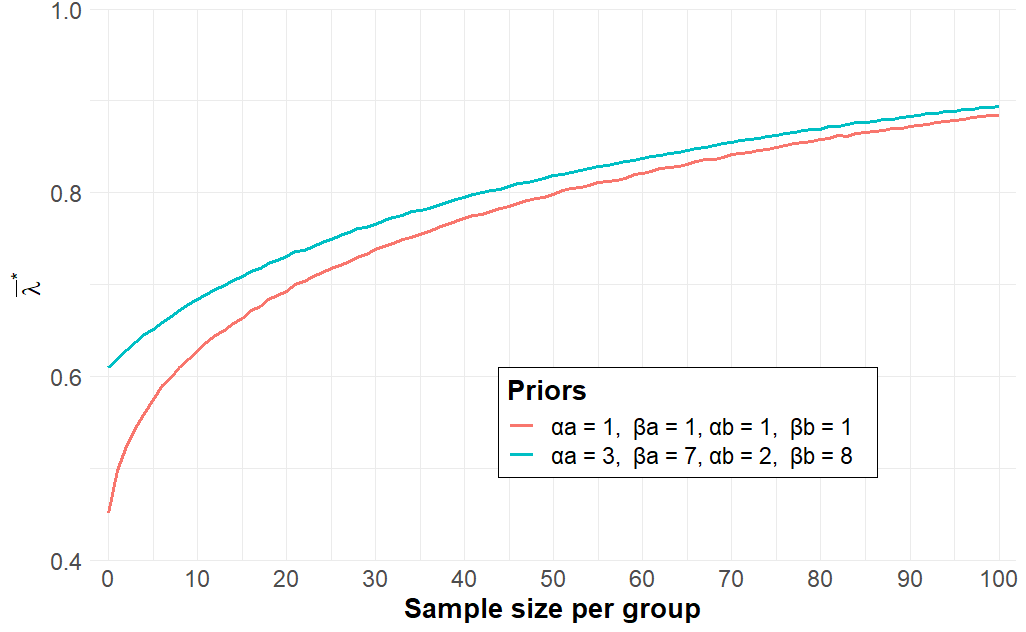}
    \label{fig:sample} %graph(100000,1,1,1,1,0.3,0.15,0.05,0.5,1)
\end{figure}

\section{SIMULATION STUDY}
We perform a simulation study to evaluate the influence of prior information on the proposed sample size calculation, based on $\lambda^*$ and $\bar{\lambda^*}$. The prior distribution was assumed to be a Beta distribution with parameters $(\alpha_i, \beta_i)$, which can be think of as information from ($\alpha_i + \beta_i$) pseudo observations with the expected response rate of $\alpha_i/(\alpha_i + \beta_i)$ . We evaluated cases where the number of pseudo observations was equivalent to 10, 20, 30, 40, and 50 patients per group, as well as the case of a vague prior, $\text{Beta}(1, 1)$.

\subsection{Evaluation of the Proposed Method with Different Response Rates}
We assessed the impact of different specifications of prior distribution on the proposed decision rule, assuming the response rates of $\tilde{\pi}_A = 0.30$ and $\tilde{\pi}_B = 0.15$. The clinically meaningful difference is set to be $d = 0.05$, with the decision parameters as $\rho = 0.5$ and $\gamma^*=0.90$. The required sample size are 39 and 65 of based on $\lambda^*$ and $\bar{\lambda}^*$, respectively. In this scenario, the evaluation criterion $\xi_{\tilde{n}_\text{min}}$ is defined as the proportion of times the treatment with a higher observed response rate is selected, with the mathematical definition given below. In our evaluation, we set the threshold $\theta=0.90$. The number of simulations was set to $m=100,000$ per scenario. It can be expressed as follows:
\begin{align*}
\xi_{\tilde{n}_\text{min}}=\frac{1}{100,000}\sum_{j=1}^{100,000} \mathbb{1}(\lambda^*_j > 0.90).
\end{align*}
Here, $\mathbb{1}(\lambda^*_j > 0.90)$ is an indicator function that takes the value 1 if $\lambda^*_j > 0.90$, and 0 otherwise. This corresponds to the proportion of times our decision rule correctly advances treatment $A$ to the confirmatory phase. Scenarios 1.1 to 1.5 in Table 3 assume a true difference between the response rates, suggesting $A$ is superior to $B$. Therefore, a higher value of $\xi_{\tilde{n}_\text{min}}$ is desirable. The sample sizes used for the evaluation were based on two different sample size approaches, corresponding to the minimum required sample sizes of 39 and 65 when the parameters were set to $\tilde{\pi}_A = 0.30$, $\tilde{\pi}_B = 0.15$, $d = 0.05$, $\rho = 0.5$, and $\gamma^* = 0.90$. Table 4 evaluates the extent to which our proposed decision rule is influenced by different levels of prior information, assuming a true difference in response rates. In Scenario 1.1, a vague prior distribution is used as the benchmark. Scenario 1.2 presents the results when the prior is perfectly consistent with the true response rates ($\pi_{\text{prior}A}=0.30, \pi_{\text{prior}B}=0.15$). Scenario 1.3 examines the results when the prior suggests some drift in the difference of response rates ($\pi_{\text{prior}A}=0.30, \pi_{\text{prior}B}=0.20$). Scenario 1.4 presents the results when the prior implies a sizable shift in the difference of response rates ($\pi_{\text{prior}A}=0.30, \pi_{\text{prior}B}=0.10$) . Finally, in Scenario 1.5, while the difference in true response rates is consistent with what the prior implies, the prior suggests considerable drift in the respective response rates ($\pi_{\text{prior}A}=0.50, \pi_{\text{prior}B}=0.35$). 

\begin{table}[hbtp]
\captionsetup{justification=centering} % Center the caption
\caption{Proportion of correct selection of treatment A when there is a true difference in response rates under various prior information}
\label{table:data_type}
\centering
\begin{tabular}{ccccccccccc}
\hline
&$\pi_A$ & $\pi_B$ & $d$ & $\rho$ & $\pi_{\text{prior}A}$ & $\pi_{\text{prior}B}$ & $\text{Beta}(\alpha_A, \beta_A)$ & $\text{Beta}(\alpha_B, \beta_B)$ & $\xi_{39}$ & $\xi_{65}$ \\
\hline
Scenario 1.1&0.30&0.15&0.05&0.5&&&(1, 1) & (1, 1) & 54.6 & 68.4\\
\hline
Scenario 1.2&0.30&0.15&0.05&0.5&0.30&0.15&(3, 7) & (2, 8) & 58.2 & 70.7\\
&&&&&&&(6, 14) & (3, 17) & 71.6 & 80.3\\
&&&&&&&(9, 21) & (5, 25) & 73.8 & 81.7\\
&&&&&&&(12, 28) & (6, 34) & 82.4 & 87.2\\
&&&&&&&(15, 35) & (8, 42) & 83.4 &88.7\\
\hline
Scenario 1.3&0.30&0.15&0.05&0.5&0.30&0.20&(3, 7) & (2, 8) & 58.2& 70.7\\
&&&&&&&(6, 14) & (4, 16) & 60.5 &73.1\\
&&&&&&&(9, 21) & (6, 24) & 63.6 &75.3\\
&&&&&&&(12, 28) & (8, 32) & 64.5 &75.8\\
&&&&&&&(15, 35) & (10, 40) & 64.9 &77.2\\
\hline
Scenario 1.4&0.30&0.15&0.05&0.5&0.30&0.10&(3, 7) & (1, 9) &69.6&78.2\\
&&&&&&&(6, 14) & (2, 18) & 81.0 &85.9\\
&&&&&&&(9, 21) & (3, 27) & 88.7 &91.0\\
&&&&&&&(12, 28) & (4, 36) & 93.2 &94.4\\
&&&&&&&(15, 35) & (5, 45) & 96.9 &97.2\\
\hline
Scenario 1.5&0.30&0.15&0.05&0.5&0.50&0.35&(5, 5) & (4, 6) & 54.7& 69.1\\
&&&&&&&(10, 10) & (7, 13) & 64.9& 76.4\\
&&&&&&&(15, 15) & (11, 19) & 64.9& 77.2\\
&&&&&&&(20, 20) & (14, 26) & 76.1& 85.3\\
&&&&&&&(25, 25) & (18, 32) & 81.0& 86.9\\
\hline
\end{tabular}
\\
  \vspace{0.5\baselineskip} % 適宜調整してください
\end{table}

Scenario 1.2 showed that increasing the prior effective sample size could results in a higher $\xi_{\tilde{n}_\text{min}}$ because of prior-data consistency. Scenario 1.3 confirmed that increasing the prior effective sample size does not increase $\xi_{\tilde{n}_\text{min}}$ when the difference in prior response rates is smaller. Scenario 1.4 was confirmed that increasing the prior effective sample size with a larger difference in response rates results in a higher $\xi_{\tilde{n}_\text{min}}$. In Scenario 1.5, though $\xi_{\tilde{n}_\text{min}}$ decreased by approximately 2\% compared to Scenario 1.2, a comparable result was still achieved.

\subsection{Evaluation of the Proposed Method with Indifferent Response Rates}
We used a different criterion than Scenario 1 to assess whether the probability of selecting treatment A would decrease, assuming the response rates are $\pi_A=0.30$ and $\pi_B=0.30$, a clinically meaningful difference of $d=0.05$ and $\rho=0.5$. The evaluation criterion for this scenario defines $\nu_{\tilde{n}_\text{min}}$ as the proportion of times selecting the more desirable group was below 90\%, i.e., $\lambda^*_j \leq \theta = 0.90$. It can be expressed as follows:
\begin{align*}
\nu_{\tilde{n}_\text{min}}=1-\xi_{\tilde{n}_\text{min}}.
\end{align*}
This corresponds to the proportion of times our decision rule considered advancing a group based on a secondary factor. In Scenario 2.1 to 2.3 in Table 4, where there is no true difference, a lower $\nu_{\tilde{n}_\text{min}}$ is preferable. As in Section 4.1, the sample sizes used for the evaluation were based on two different sample size approaches, corresponding to the minimum required sample sizes of 39 and 65 when the parameters were set to $\tilde{\pi}_A = 0.30$, $\tilde{\pi}_B = 0.15$, $d = 0.05$, $\rho = 0.5$, and $\gamma^* = 0.90$. Table 4 evaluates how our proposed decision rule is influenced by prior data when there is no true difference in response rates. In Scenario 2.1, a vague prior distribution is assumed as the benchmark. Scenario 2.2 presents when prior suggests a difference of 0.10 in response rates ($\pi_{\text{prior}A}=0.30$, $\pi_{\text{prior}B}=0.20$). Scenario 2.3 presents the results when prior suggests a difference of 0.20 in response rates ($\pi_{\text{prior}A}=0.30$, $\pi_{\text{prior}B}=0.10$). 
\begin{table}[hbtp]
\captionsetup{justification=centering} % Center the caption
\caption{Proportion of decisions considering a secondary factor when there is no true difference in response rates under various prior information}
\label{table:data_type}
\centering
\begin{tabular}{ccccccccccc}
\hline
&$\pi_A$ & $\pi_B$ & $d$ & $\rho$ & $\pi_{\text{prior}A}$ & $\pi_{\text{prior}B}$ & $\text{Beta}(\alpha_A, \beta_A)$ & $\text{Beta}(\alpha_B, \beta_B)$ & $\nu_{39}$ & $\nu_{65}$ \\
\hline
Scenario 2.1&0.30&0.30&0.05&0.5&&&(1, 1) & (1, 1) & 92.4 & 93.9\\
\hline
Scenario 2.2&0.30&0.30&0.05&0.5&0.30&0.20&(3, 7) & (2, 8) & 91.2 & 92.4\\
&&&&&&&(6, 14) & (4, 16) & 90.9 &92.2\\
&&&&&&&(9, 21) & (6, 24) & 89.6 &91.5\\
&&&&&&&(12, 28) & (8, 32) & 87.3 &90.0\\
&&&&&&&(15, 35) & (10, 40) & 86.8 &89.2\\
\hline
Scenario 2.3&0.30&0.30&0.05&0.5&0.30&0.10&(5, 5) & (4, 6) & 86.4& 89.2\\
&&&&&&&(6, 14) & (2, 18) & 79.1 &84.8\\
&&&&&&&(9, 21) & (3, 27) & 67.1 &77.6\\
&&&&&&&(12, 28) & (4, 36) & 55.0 &68.6\\
&&&&&&&(15, 35) & (5, 45) & 44.7 &61.0\\
\hline
\end{tabular}
\\
  \vspace{0.5\baselineskip} % 適宜調整してください
\end{table}

Scenario 2.2 illustrates when there is no true difference, yet the investigator assumed a difference and sets a misleading prior distribution. However, regardless of whether $\nu_{39}$ or $\nu_{65}$ is considered, when prior data are fewer than observed data, the proportion of simulations in which the probability of selecting the more desirable group was below 90\% remained at approximately 90\%. This suggests that even when incorrect prior information is used, the probability of making a decision based on efficacy remains very low.Scenario 2.3 illustrates a case in which there is no true difference, yet the researcher incorrectly assumes a difference and overestimates the treatment effect size compared to Scenario 2.2. As the assumed difference between $\pi_{\text{prior A}}$ and $\pi_{\text{prior B}}$ increases and a stronger prior is incorporated, the proportion of correctly selecting the more desirable group decreases below 90\%. In other words, the probability of incorrectly selecting the more desirable group increases.

\section{REAL DATA ANALYSIS}
We revisit a phase II RCT \cite{wild} in elderly patients with HER2-positive metastatic breast cancer , and analyse the data retrospectively using the proposed approach. This clinical trial, originally planned based on the SG design, investigates a metronomic chemotherapy with pertuzumab and trastuzumab. The primary endpoint for the two treatment groups, pertuzumab plus trastuzumab (PT) and PT plus metronomic chemotherapy (PTM), was the progression-free survival rate at six months. The expected progression-free survival rates were $\pi_{\text{PT}}=0.40$ and $\pi_{\text{PTM}}=0.55$, respectively, with a clinically meaningful difference $d$ of 0.10. When the required sample size was 40 per group, which was the same as in the original trial, the value of $\lambda = P_{Corr} + \frac{1}{2} P_{Amb}$ was 0.81.

Table 5 shows the results of $\lambda^*$ for required sample size of 40 per group in the SG design (Frequentist framework) and the Bayesian framework. Details about the SG design can be found in the Appendix.
\begin{table}[hbtp]
\caption{Comparison between the SG design and our Bayesian version in terms of  $\lambda^\ast$}
\label{table:data_type}
\centering
\begin{tabular}{cccccc}
\hline
SG design & \multicolumn{2}{c}{$\pi_{\text{PTM}}$}& \multicolumn{2}{c}{$\pi_{\text{PT}}$} &$\lambda$\\
\hline
&\multicolumn{2}{c}{0.55}&\multicolumn{2}{c}{0.40}&0.81\\
\hline
\hline
Bayesian SG design&$\pi_{\text{PTM}}$&prior distribution&$\pi_{\text{PT}}$&prior distribution&$\lambda^*$\\
\hline
&0.55&$\pi_{\text{PTM}} \sim \text{Beta}(1, 1)$&0.40&$\pi_{\text{PT}} \sim \text{Beta}(1, 1)$&0.82\\
&0.55&$\pi_{\text{PTM}} \sim \text{Beta}(1, 1)$&0.40&$\pi_{\text{PT}} \sim \text{Beta}(26, 40)$&0.86\\
\hline
\end{tabular}
\end{table}

When both $\pi_{\text{PTM}}$ and $\pi_{\text{PT}}$ have a Beta(1, 1) prior each, our proposed design yields $\lambda^*=0.82$, which is close to the value of $\lambda$ in the frequency calculation. In practical clinical trials, sample size calculations may be based on assumed true progression-free survival rates for each group, derived from previous research results. When incorporating such information into the prior distribution for progression-free survival rates, the calculated $\lambda^*$ increases to 0.86.

\begin{table}[hbtp]
\caption{Sample size calculation using the exsisting method and the proposed method when $\lambda=\lambda^*=0.80$}
\label{table:data_type}
\centering
\begin{tabular}{cccccc}
\hline
SG design & \multicolumn{2}{c}{$\pi_{\text{PTM}}$}& \multicolumn{2}{c}{$\pi_{\text{PT}}$} &required sample size per group\\
\hline
&\multicolumn{2}{c}{0.55}&\multicolumn{2}{c}{0.40}&40\\
\hline
\hline
Bayesian SG design&$\pi_{\text{PTM}}$&prior distribution&$\pi_{\text{PT}}$&prior distribution&required sample size per group\\
\hline
&0.55&$\pi_{\text{PTM}} \sim \text{Beta}(1, 1)$&0.40&$\pi_{\text{PT}} \sim \text{Beta}(1, 1)$&40\\
&0.55&$\pi_{\text{PTM}} \sim \text{Beta}(1, 1)$&0.40&$\pi_{\text{PT}} \sim \text{Beta}(26, 40)$&20\\
\hline
\end{tabular}
\end{table}

Table 6 reports the sample size required using the existing method and the proposed method. In the frequentist calculation, the threshold value of $\lambda$ is used to calculate the required sample size. When a more informative Beta(26, 40) prior is placed on $\pi_{\text{PT}}$, each group requires 20 sample size. The results from Tables 5 and 6 confirm a reduction in the required number of cases when prior information is considered.

\section{SHINY APPLICATION}
To achieve our goal of making Bayesian designs more accessible to clinicians, we developed a user-friendly interface using R Shiny , at \url{https://mokakomaki.shinyapps.io/my_shiny_app/}. Figures 3 and 4 show the actual Shiny application. 
\\

The strengths of this software application are as follows:\\
1. By entering the required parameters, the necessary sample size calculation results are output within seconds.\\
2. No statistical programming skills are required.\\
3. Text for protocols and Statistical Analysis Plans (SAPs) is automatically generated.\\
4. It is easy to compute statistical analysis results after the trial ends.\\
5. The simple design ensures readability.\\
\\
%\texttt{https://mokakomaki.shinyapps.io/my_shiny_app/}

\begin{figure}[h!]
    \centering
    % 1枚目の画像 (a)
    \begin{minipage}{0.48\textwidth}
        \centering
        \includegraphics[width=\textwidth]{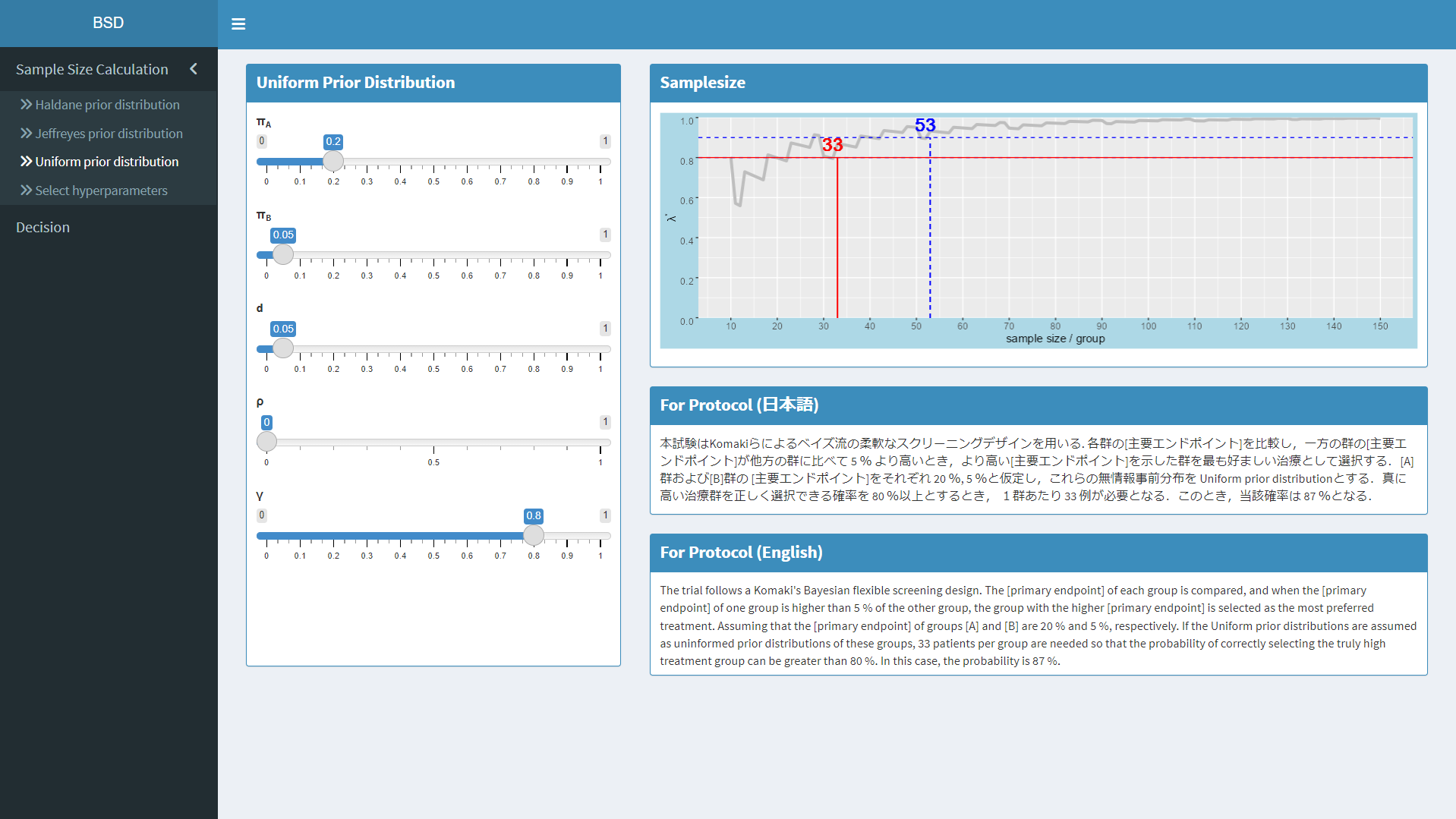}
        \\(a) Main interface
    \end{minipage}
    \hfill
    % 2枚目の画像 (b)
    \begin{minipage}{0.48\textwidth}
        \centering
        \includegraphics[width=\textwidth]{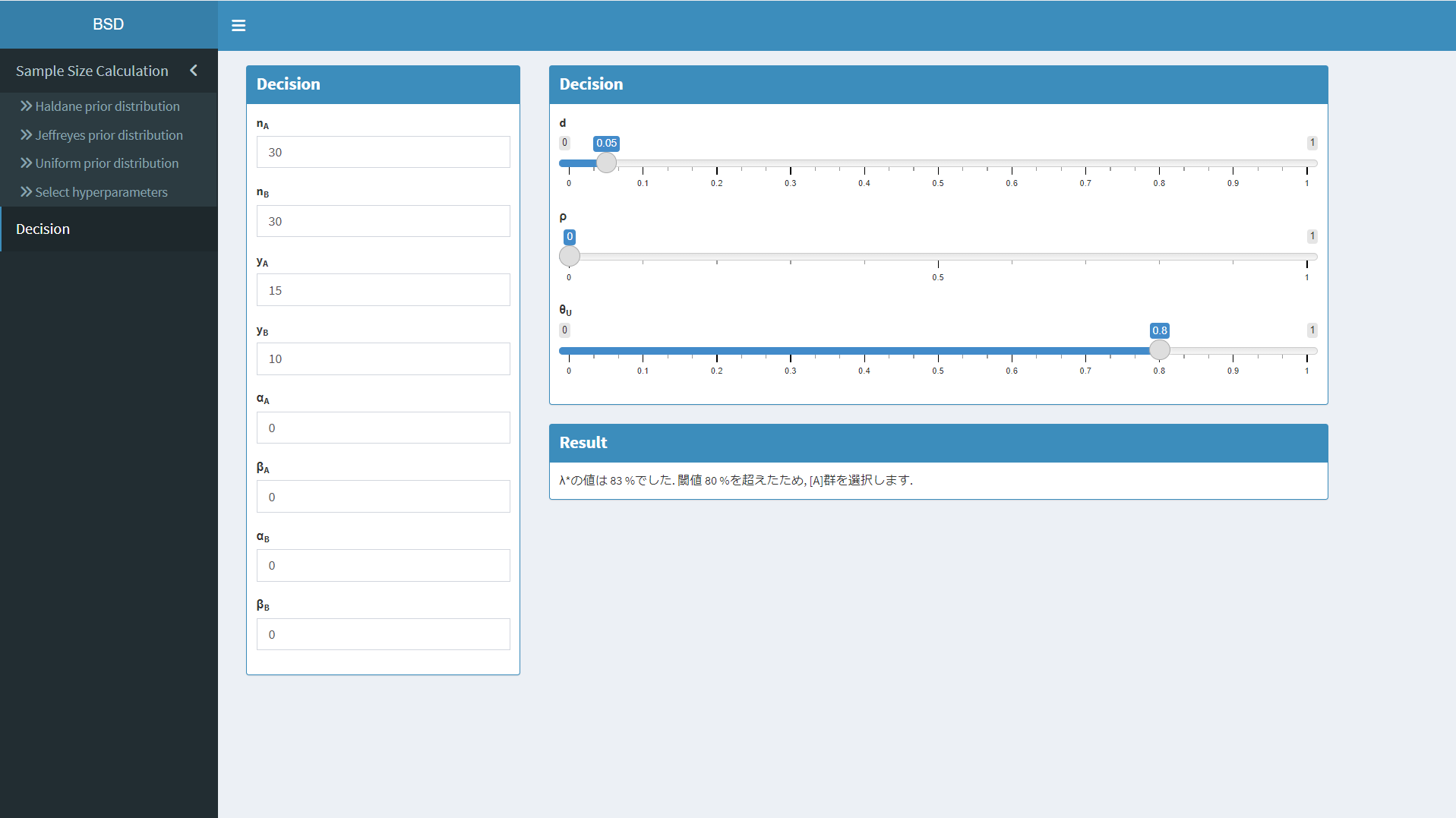}
        \\(b) Analysis results
    \end{minipage}
    \caption{Interface of our R Shiny that implements the proposed Bayesian design.}
    \label{fig:shiny_interface}
\end{figure}

\section{DISCUSSION}
Our proposed Bayesian design is motivated by the SG design, which incorporates pragmatic considerations from clinical practice. That is, response rate to a treatment is often only one of many considerations in determining the recommendation for further investigation and ultimately to patients \cite{sar}. However, it is important to note that the decision rule in our proposed design differs slightly from that of the SG design. In our approach, treatment selection is based on efficacy, as determined by whether the probability of correct selection or the threshold of $\lambda^*$ is exceeded. This methodology is directly linked to sample size calculation, resulting in a more straightforward and interpretable design.

Our proposed design can be implemented by two ways of sample size calculation. Choice between these methods depends on the purpose of the clinical trial. The method based on $\lambda^*$ is considered to improve the overall success rate of the trial. In this work, we focused on the method based on $\lambda^*$, which is expected to be commonly used. As shown in Figure 1, the method based on $\lambda^*$ shows that $\lambda^*$ becomes unstable when the sample size is small. To address this, we proposed sample size calculations based on $\bar{\lambda^*}$, which were found to be more stable even with small sample sizes. This method tends to result in larger required sample sizes compared to the $\lambda^*$-based method, which is unsurprising for additional randomness around the parameters accommodated. On the other hand, particularly when $\rho=\frac{1}{2}$ and $\gamma^*=0.80$, there are cases where the required sample sizes based on $\lambda^*$ and $\bar{\lambda^*}$ differ by only a little. It is advisable to calculate both sample sizes based on $\lambda^* $and $\bar{\lambda^*}$ in the trial planning. If a resulting sample size is feasible, the calculation using $\bar{\lambda^*}$ is recommended. In addition, in the present work, vague prior distributions were employed for the calculation of the required sample size. The necessity for a larger sample size may be reduced if reliable prior information from previously conducted studies can be integrated. When incorporating prior information, it is of the utmost importance to carefully consider whether the prior information is compatible with the trial data. We refer the interested readers to Zheng et al. (2023a) for robust Bayesian sample size determination by using mixture priors and a discounting parameter \cite{HZ}.

The evaluation was conducted using clinically plausible scenarios in the simulation study. Table 3 assessed how the results are affected when prior data is introduced in cases where the response rates differ by some extent. Incorporating prior data with the same response rates as the true rates increased the probability of selecting treatment A compared to when vague priors were used. Moreover, even when prior data with different response rates but the same difference in response rates between the groups were used, the probability of selecting treatment A increased without substantially altering the decision. These findings suggest that incorporating appropriate prior information increases the likelihood of selecting the desired group when there is a true difference in response rates. Evaluations were also conducted for cases where the difference in prior response rates was smaller or larger than the true response rate difference, and it was found that the probability of selecting treatment A increased compared to the case of vague priors. The simulation suggests that when the relationship between response rates for both treatments is consistent between the prior and the data, the likelihood of selecting the better group is higher. Table 4 assessed how the results are influenced when prior data is introduced in cases where there is no true difference in response rates. When the difference between the true response rates and the prior response rates was around 10\%, the impact on the results was not significant. However, when the amount of prior data exceeded that of the observed data, the prior data had a stronger impact on the results, so caution is needed in determining how much prior data to incorporate. We note that dynamically handling prior-data conflict in early phase clinical trials\cite{HZ_1} is beyond the scope of this paper. This could be a research avenue to pursue in the future.

The definition of the second factor, which forms the basis of our proposed Bayesian design as well as the SG design, is not strictly fixed. For example, in cases where no difference is observed between the response rates, the decision to select a treatment for the confirmatory phase III clinical trial may be based on clinical judgment, which raises concerns about potentially arbitrary outcomes. In the field of oncology, safety is likely the second most important factor after efficacy, and it is desirable to define and quantify safety as the second factor. On the horizon of precision medicine, we are interested in extending our Bayesian design to enable simultaneous selection of treatment in multiple patient subgroups \cite{HZ_2}. There may also be a need for our design to remain exploratory, flexible, and adaptable.

In conclusion, this paper proposed a Bayesian treatment selection design as an analogue of that proposed by Sargent and Goldberg. The Bayesian version enables incorporation of external-trial information into the prior distributions, which can mean a reduction in the required sample size. The associated sample size calculation can be performed using our easy-to-implement web application. Our proposed methodology can be extended to involve sequential decision making as well as to consider heterogeneous patient populations.

\newpage
\noindent
\section*{APPENDIX}
\subsection*{A quick primer on the Sargent and Goldberg's Design }
We summarise the Sargent and Goldberg's design here for interested readers' information: Consider a randomised Phase II trial with two treatment groups ($A, B$). Let $\pi_A$ denote the true response rate on the better treatment (without loss of generality
treatment $A$), $\pi_B$ denote the true response rate on the poor treatment (treatment $B$), define $\delta = \pi_A - \pi_B$, and let $n_A$ and $n_B$ be the number of patients per treatment. Let $p_A$ and $p_B$ denote the observed response rates on treatments $A$ and $B$, respectively. Taking the primary endpoint as the binary variable response rate, with $d$ representing a clinically meaningful difference greater than 0. If the observed difference in response rates between groups $A$ and $B$ is greater than the predetermined $d$, indicating a substantial difference, the superior group is selected. However, if the observed difference falls within $\pm \ d$ of the predetermined value, indicating no discernible difference, considerations of factors other than the response rate come into play.

The sample size calculation is outlined as follows: Assume true response rates for groups $A$ and $B$, and define the following as the probabilities of correct selection and ambiguity:
\begin{align*}
P_{Corr} &= \mathrm{Pr}[p_A-p_B>d \ | \ \pi_A, \pi_B],\\
P_{Amb} &= \mathrm{Pr}[-d \leq p_A-p_B \leq d \ | \ \pi_A, \pi_B],
\end{align*}
Assuming that we take a number $n_i \ (i=A, B)$ of patients from each treatment, we can see that these two probabilities are actually equal to
\begin{align*}
P_{Corr} &= \sum_{x_A=0}^{n_A}\sum_{x_B=0}^{n_B}1_{\{(x_A-x_B)/n>d\}}\begin{pmatrix}n_A\\x_A\end{pmatrix} \begin{pmatrix}n_B\\x_B\end{pmatrix} \pi_A^{x_A}(1-\pi_A)^{n_A-x_A}\pi_B^{x_B}(1-\pi_B)^{n_B-x_B},
\end{align*}
where $1_{inequality}$ is theindicator function, and equals 1 if thecondition of theinequality holds,
0 otherwise, and
\begin{align*}
P_{Amb} &= \sum_{x_A=0}^{n_A}\sum_{x_B=0}^{n_B}1_{\{-d \leq (x_A-x_B)/n \leq d\}}\begin{pmatrix}n_A\\x_A\end{pmatrix} \begin{pmatrix}n_B\\x_B\end{pmatrix} \pi_A^{x_A}(1-\pi_A)^{n_A-x_A}\pi_B^{x_B}(1-\pi_B)^{n_B-x_B},
\end{align*}
Note that as $n_A$ and $n_B$ are large, we can approximate the above two probabilities by using the central limit theorem. That is, the probability of correct selection
\begin{align*}
P_{Corr} \dot{=} \ \mathrm{Pr}\left[\frac{Z>(d-\delta)}{\sqrt{\text{var}}}\right],
\end{align*}
where $Z$ denotes the standard normal random variable, $\text{var}=[\pi_A(1-\pi_A)+\pi_B(1-\pi_B)]/n$ and $n=n_A=n_B$.

Similarly, we can show that the ambiguity probabilities
\begin{align*}
P_{Amb} \dot{=} \ \mathrm{Pr}\left[ \frac{Z\leq(d-\delta)}{\sqrt{\text{var}}}\right]-\mathrm{Pr}\left[ \frac{Z\leq(-d-\delta)}{\sqrt{\text{var}}}\right],
\end{align*}
Furthermore, set a predetermined threshold, $\gamma$. The parameter $\lambda$ represents a value that considers both the probability of correct selection and the probability of ambiguity and is defined as:
\begin{align*}
\lambda = P_{Corr} + \rho P_{Amb},
\end{align*}
Here, $\rho$ denotes the extent to which ambiguity is considered and typically takes values of 0 or $\frac{1}{2}$. The required sample size for this design is determined when
\begin{align*}
\lambda > \gamma,
\end{align*}
By incorporating the probability of ambiguity, which considers secondary factors, this design allows for a reduction in the required sample size. Please refer to the more details within the reference \cite{let}.

%$P_{Corr}$ and $P_{Amb}$ in Equations (1) and (2) were assumed, $\lambda$ was calculated, and subsequently, the required sample size was determined. On the other hand, the number of cases in square brackets [] in Table 4 was calculated using the normal approximation method based on the central limit theorem. $P_{Corr}$ and $P_{Amb}$ in Equations (3) and (4) were assumed, respectively, and the required number of cases was obtained after calculating $\lambda$.

%\section{ランダム化選択デザイン}
%\quad ランダム化選択デザインの1つであるSargentらが提案したデザインは, 第III相臨床試験に進める推奨治療を決定する際に, 主要評価項目のみで推奨治療を決定しきれないときに他の要因も考慮できる柔軟なデザインである. 他の要因は, 毒性やコスト, 投与の容易さ, 生活の質など患者さんにとって第二の重要な要因を指している. デザインの概要は以下のとおりである. 2群 (A, B)のランダム化第二相試験を想定する. 主要評価項目を二値変数の奏効割合として, $d$は0よりも大きい臨床的に意義のある差とする. A群とB群の観測された奏効割合の差が事前に決められていた$d$よりも大きい場合, 2群間に十分な差が認められたため, より優れた群を選択する. 一方で, A群とB群の観測された奏効割合の差が事前に決められていた$\pm d$以下だったとき, 2群間に差が認められないと判断し奏効割合以外の他の要因を考慮する. 
%症例数設計は以下のとおりである. あらかじめA群とB群の真の奏効割合を仮定し, 正選択確率と曖昧確率として以下を定義する.
%\begin{align*}
%P_{Corr} = \mathrm{Pr} [ \text{A群の推定値} - \text{B群の推定値} > d ]
%\end{align*}
%\begin{align*}
%P_{Amb} = \mathrm{Pr}[ -d < \text{A群の推定値} - \text{B群の推定値} < d ]
%\end{align*}
%さらに, 事前設定された閾値を$\gamma$とする.\\
%$\lambda$は正選択確率と曖昧確率を考慮した値であり, 以下のように定義される.
%\begin{align*}
%\lambda = P_{Corr}+\rho P_{Amb}
%\end{align*}
%$\rho$は曖昧確率を含む程度を表しており, 通常0か0.5を使用する.
%以上の定義より
%\begin{align*}
%\lambda > \gamma
%\end{align*}
%となる症例数をこのデザインの必要症例数としている.
%このように第二の他の要因を考慮する曖昧確率を含むことで, 症例数を抑えることができるような設計となっている. 

%%%%%%%%%%%%%%%%%%%%%%%%%Appendixで使うかも%%%%%%%%%%%%%%%%%%%%%%%%%

\end{document}